\documentclass[prb, twocolumn, amsmath, amssymb]{revtex4}

\setcitestyle{numbers,square}
\usepackage{amsmath,amssymb}
\usepackage{tabularx}
\usepackage{graphicx}
\usepackage{bmpsize}
\usepackage{bm}
\usepackage{color}
\usepackage{amssymb}
\usepackage[version=3]{mhchem}
\usepackage{yhmath}
\DeclareMathAlphabet{\mathpzc}{OT1}{pzc}{m}{it}
\newcommand{\nn}{\nonumber}

\def\diag{\mathrm {diag}}

\begin{document}
\makeatletter
\renewcommand\@biblabel[1]{[#1]}
\makeatother

\preprint{APS/123-QED}

\title{Topological Domain Walls in Graphene Nanoribbons with Carrier Doping}

\author{Takuto Kawakami}
\email{t.kawakami@pq.phys.sci.osaka-u.ac.jp}
\affiliation{Department of Physics, Osaka University, Toyonaka, Osaka 560-0043, Japan}
\author{Gen Tamaki}
\affiliation{Department of Physics, Osaka University, Toyonaka, Osaka 560-0043, Japan}
\author{Mikito Koshino}
\affiliation{Department of Physics, Osaka University, Toyonaka, Osaka 560-0043, Japan}

\date{\today}

\begin{abstract}

We theoretically study magnetic ground states of doped zigzag graphene nanoribbons and the emergence of topological domain walls.
Using the Hartree-Fock mean-field approach and an effective continuum model, 
we demonstrated that the carrier doping stabilizes a magnetic structure with alternating antiferromagnetic domains, 
where the doped carriers are accommodated in topological bound states localized at the domain wall.
The energy spectrum exhibits a Hofstadter-like fractal spectral evolution as a function of the carrier density,
where minigaps are characterized by the Chern number associated with the adiabatic charge pump in moving domain walls. 
A systematic analysis for nanoribbons with different widths revealed that the ferromagnetic domain-wall phase emerges in relatively wide ribbons, while the colinear domain-wall phase arises in narrower ribbons.


\end{abstract}


\maketitle

\section{Introduction}

Zigzag graphene nanoribbon (ZGNR) and its exotic physics properties have been extensively studied in the past couple of decades~\cite{netoreview,schwierz2010, dassarma2011review,yazyevreview,meunier2016review, celis2016review}. 
One of the most prominent features of ZGNR is the emergence of edge states, 
which form one-dimensional flat bands at the charge neutral point~\cite{nakada1996,fujita1996}.
In the presence of electron-electron interaction, the large density of states of the flat bands 
leads to occurrence of an antiferromagnetic (AFM) order, where spins at the two edges are polarized in an antiparallel manner,
and an energy gap opens between the flat bands
~\cite{fujita1996, son2006,sonprl2006,pisani2007,rossier2007,feldner2011,dutta2012}. 
The one-dimensional edge magnetism of ZGNR is expected to provide an ideal platform for spintronics devices~\cite{yazyev2008,ma2011,cui2016,sancho2017,sanz2022,pizzochero2022}.
In experiments, GNRs with atomically precise edge structures have been fabricated by
the state-of-the-art techniques~\cite{saraswat2021}, such as STM nanolithography~\cite{tapaszto2008,magda2014}, 
unzipping carbon nanotubes~\cite{kosynkin2009,jiao2009,jiao2010,kosynkin2011},  
electron-beam lithography~\cite{wu2018}, 
and templated bottom-up synthesis using appropriate monomers~\cite{cai2010,narita2014,han2014,ruffieux2016,wang2016,kawai2015, rizzo2018,groning2018,kitao2020,kolmer2020}.
The AFM-induced band gap was actually observed in precisely-fabricated ZGNRs, by using tunnelling spectroscopy\cite{ritter2009,tao2011,magda2014} and point contact spectroscopy~\cite{blackwell2021}.

The thermodynamic stability of the magnetic order in ZGNR is supported by the energy gap between the flat bands opened by the AFM exchange field~\cite{nakada1996,fujita1996,son2006,sonprl2006}.
Here we ask how the magnetic order survives 
when a ZGNR is carrier-doped by applying gate voltage or chemical doping.
Naively it seems that doping destabilizes the AFM order, as the Fermi energy is shifted to the band edges with large densty of states above or below the energy gap.
The system is expected to take some different ordered states to gap out the spectrum
right at the Fermi energy.

\begin{figure}[b]
\begin{center}
	\includegraphics[width=85mm]{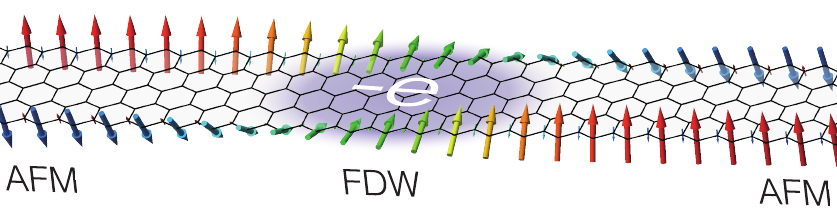}
	\caption{Schematic picture of the typical domain wall in doped ZGNR.
	Arrows indicate the spin density localized at the zigzag edge.
	Doped carrier is trapped at the domain wall.
	}\label{fig:schem}
\end{center}
\end{figure}

In this paper, we study the magnetic and electronic properties of ZGNRs with various ribbon widths in a wide range of carrier doping,
on the basis of the Hartree-Fock mean-field framework.
In the low carrier density regime, we observe that a magnetic domain structure with alternating AFM order (Fig.~\ref{fig:schem}) is spontaneously formed, consistently with Ref.~\cite{sancho2017},
where the doped carriers are accommodated in topological localized states bound to the domain walls.
We find that the domain-wall bound state is topologically protected and it can be described by a Jackiw-Rebbi type model~\cite{jackiw} with a mass inversion.
When the carrier density is further increased, the distance of neighboring domain walls becomes smaller, and
the magnetic domain structure eventually crosses over to the spin and charge density wave.
The electronic spectrum as a function of the electron density exhibits a fractal pattern
similar to the Aubry-Andr\'e model~\cite{lang2012, kraus2012, kraus2012_2} and the Hofstadter butterfly~\cite{hofstadter},
due to a competition of the periodic magnetic structure and the atomic lattice constant.
We show that each gap in the spectrum can be characterized by a Chern number, which is associated with
the quantized charge pump under an adiabatic parallel shift of the periodic domain walls.

\begin{figure*}[t]
\begin{center}
	\includegraphics[width=170mm]{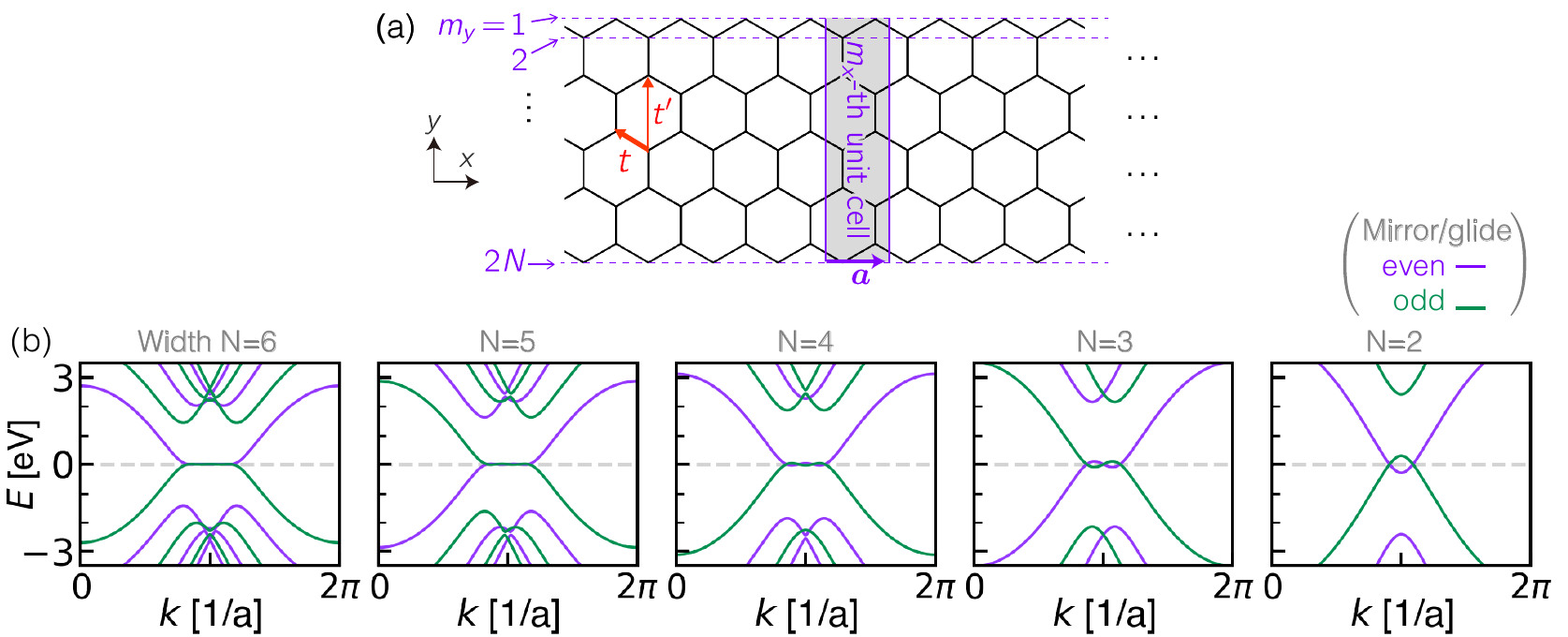}
	\caption{(a) Lattice structure of width $N=5$ ZGNR and its non-interacting dispersion relation.
	Shaded region indicates the ribbon unit cell. $\bm{a}$ is lattice translation vector.
	$m_x$ and $m_y$ is a set of labels characterizing each lattice points.
	To discuss the domain wall structures we calculate the long supercell along $x$-direction.
    (b) Dispersion relations for ZGNR with general width $N$.
    Purple and green curves in $N=6$, $4$, and $2$ ($N=5$, $3$) are for the mirror even and odd (glide even and odd) states (see Appendix~\ref{sec:non-interact} for more details).
	}\label{fig:honeycomb}
\end{center}
\end{figure*}

This paper is organized as follows. 
In Sec.~\ref{sec:formulation}, we introduce the mean-field framework used in this paper. 
In Sec.~\ref{sec:tdw}, we calculate the spin and charge density structure 
in the ground state of the carrier doped ZGNR. 
We also discuss the electronic spectral evolution as a function of the carrier density.
In Sec.~\ref{sec:top} we discuss about the topological origin of the domain wall bound state 
using the winding number and Jackiw-Rebbi arguments in the low-energy continuum model.
In Sec.~\ref{sec:stability} we evaluate the thermodynamical stability of the domain walls, and obtain a phase diagram of doped ZGNR by considering finite temperature effect. 
In Sec.~\ref{sec:stability}, we present similar analyses for ZGNRs with various widths.
Finally, a brief summary is given in Sec.~\ref{sec:summary}

\section{Hartree-Fock Hubbard Model} \label{sec:formulation}
We define the structure of ZGNR of the width $N$ as illustrated in Fig.~\ref{fig:honeycomb}(a). 
The ribbon is extended along $x$ direction, and a unit cell (gray square) includes $2N$ atoms.
We model electrons in the system by a tight-binding model,
\begin{gather}\label{eq:hamilt}
	H = H_0 + H_{\mathrm{int}} \\
	{H}_0 = \sum_{ij} t^{ij} \Psi^\dag_i \Psi_j \label{eq:h0}\\
	H_{\mathrm{int}} = U \sum_j \psi_{j\uparrow}^\dag\psi_{j\uparrow}\psi_{j\downarrow}^\dag \psi_{j\downarrow} \label{eq:hint},
\end{gather}
where ${\psi}_{js}$ is the field operator at site $j$ with spin $s=\uparrow, \downarrow$
and $\Psi^\dag_j= [{\psi}_{j\uparrow }^\dag, {\psi}_{j\downarrow}^\dag]$.
$H_0$ is the non-interacting Hamiltonian, where
we take into account the nearest-neighbor hopping $t\approx 3.0$ eV and the third  nearest-neighbor hopping $t'\approx0.29$ eV,
as illustrated in Fig.~\ref{fig:honeycomb}(a).
The $t'$ is responsible for the band overlapping at charge neutral point in narrow ZGNRs~\cite{kivelson1983}.
Diagonalizing $H_0$, we obtain the non-interacting band structure of the ZGNR as shown in Fig.~\ref{fig:honeycomb}(b).
In increasing $N$, 
we see that flat bands of the edge-localized modes extend at $E=0$. ~\cite{fujita1996,nakada1996}

$H_{\rm int}$ is the electron-electron interaction part, where we only take on-site Coulomb repulsion with coupling constant $U\approx3.4$ eV. 
Atomic sites are labeled by index $j=(m_x,m_y)$ with $m_x=1,2,\cdots$ for the ribbon unit cell and $m_y=1,2,\cdots,2N$ for $y$-coordinate [see Fig.~\ref{fig:honeycomb}(a)].
Hereafter $N$ is referred to as the width of the ribbon.

We employ the mean-field approximation for $H_{\rm int}$ as,
\begin{align}
	H_{\mathrm{int}} &\approx H_{\mathrm{HF}}
	\nn\\
	 &= \frac{U}{2}\sum_{j,s}\Big( \langle\psi_{j,s}^\dag \psi_{j,s}\rangle \psi_{j,-s}^\dag\psi_{j,-s}  
	 \!-\! \frac{1}{2}\langle\psi_{j,s}^\dag \psi_{j,s}\rangle \langle\psi_{j,-s}^\dag\psi_{j,-s}\rangle \nn\\
	 &\quad -\! \langle\psi_{j,s}^\dag \psi_{j,-s}\rangle \psi_{j,-s}^\dag\psi_{j,s} 
	 \! +\! \frac{1}{2}\langle\psi_{j,s}^\dag \psi_{j,-s}\rangle \langle\psi_{j,-s}^\dag\psi_{j,s}\rangle \Big) \nn \\
	&= -\frac{U}{4} \sum_{j} \Psi_j^\dag  \big[ n^\mu(\bm{r}_j) \sigma_\mu\big] \Psi_j + E_{\mathrm{C}}, 
\end{align}
where $n^{\mu}(\bm{r}_j)$ is the mean field and 
$E_{\mathrm{C}}$ is a constant energy shift given by
\begin{gather}
	n^{\mu}(\bm{r}_j) = \langle \Psi_j^\dag \sigma^\mu \Psi_j \rangle, \label{eq:sc}\\
	E_{\mathrm{C}} = \frac{U}{8} \sum_{j} n^\mu(\bm{r}_j) n_\mu(\bm{r}_j).\label{eq:Ec}
\end{gather}
Here $\sigma^\mu=(\sigma_0, \bm{\sigma})$ and  $\sigma_\mu=(-\sigma_0, \bm{\sigma})$ are 
four vectors composed of the identity matrix $\sigma_0$ and Pauli matrices $\bm{\sigma}=(\sigma_x,\sigma_y, \sigma_z)$, 
and $\bm{r}_j=(x_j,y_j)$ indicates the position of the site $j$.

We assume that the mean field is periodic in $x$-direction with a period of $qa$  (i.e., $q$ ribbon unit cells), and apply the Bloch theorem for the supercell.
Here $q$ is a parameter and it should be determined to minimize the free energy under a given electron density.
We expect that the number of doped electrons per a supercell is an integer $M$ to make the Fermi energy come to a Bragg gap of the superlattice. Under this condition, the superlattice period $q$ and the number of electrons per site, $\bar{n}$, are related by $\bar{n} = M/(2Nq)$, where $2Nq$ is the total number of sites in a supercell.
In our mean field calculation, we find that the optimized structure has a period of $M=2$ (i.e., two electrons per a supercell), and therefore $q= 1/(N\bar{n})$.

A single-particle Schr\"odinger equation including the periodic mean field is written as
\begin{gather}
	{\sum_{j}} \left\{ H_{k}^{ij}\!-\!\frac{U}{4} n^{\mu}(\bm{r}_j)\sigma_{\mu}\delta^{ij} \right\}u_{\nu k}(\bm{r}_j) = E_{\nu k} u_{\nu k}(\bm{r}_i), \label{eq:hfeq}
\end{gather}
where $k$ is the Bloch wave number defined in the superlattice Brillouin zone, $-G/2 \leq k \leq G/2$, with $G=2\pi/qa$,
 $u_{\nu k}(\bm {r}_j)=[u_{\nu k \uparrow}(\bm{r}_j), u_{\nu k \downarrow}(\bm{r}_j)]^T$ is the periodic part of the Bloch wave function of the band index $\nu$,
and we defined  $H^{ij}_k = t^{ij} {e^{ik(x_i-x_j)}}$. 

Using the obtained wave function, self-consistent mean-field Eq.~\eqref{eq:sc} is expressed as
\begin{gather}
	n^{\mu}(\bm{r}_j) =  \sum_{\nu} \int \frac{dk}{|G|} u_{\nu k}^\dag(\bm{r}_j) \sigma^{\mu} u_{\nu k} (\bm{r}_j)f(E_{\nu k}) \label{eq:mf}
\end{gather}
where  $f(E) = [1+e^{(E-E_{\mathrm{F}})/k_{\mathrm{B}}T}]^{-1}$ is the Fermi distribution function, and
the integral in $k$ is taken over the superlattice Brillouin zone.
The Fermi energy $E_{\mathrm{F}}$ is introduced to fix the doped carrier per a site,
\begin{equation}\label{eq:nbar}
\bar{n}=\frac{1}{2 Nq} \sideset{}{'}\sum_{j} [n(\bm{r}_j)-1].
\end{equation}
Here the primed sum runs over the sites $j$ in the super unit cell.

\begin{figure}[t]
\begin{center}
	\includegraphics[width=85mm]{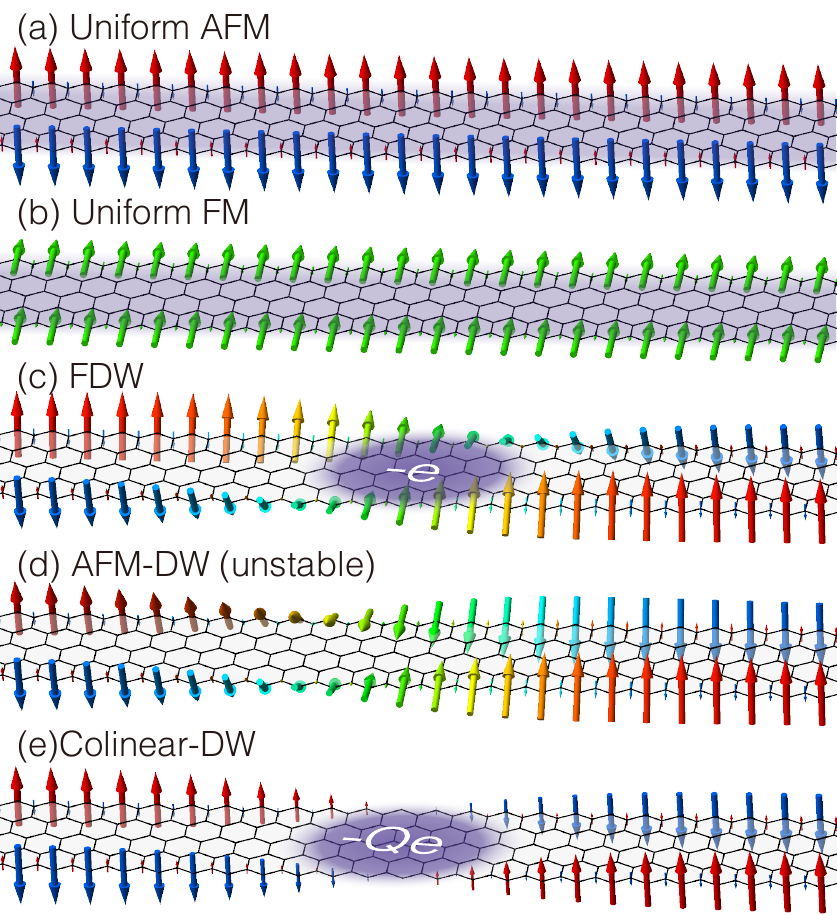}
	\caption{Candidates of stable orders. Arrows indicates the direction of spin and shaded areas schematically illustrates the charge distribution. (See text for details.)
	}\label{fig:sols}
\end{center}
\end{figure}

We solve Eqs.~\eqref{eq:hfeq}, \eqref{eq:mf}, and \eqref{eq:nbar}
by numerical iterations 
starting from different initial conditions.
We obtain several types of ordered structures listed in Fig.~\ref{fig:sols}.
For each of these states, we calculate the Helmholtz free energy density per a single carbon site~\cite{fetter, thoulessbook, kitabook}, 
\begin{gather} 
	F= -\frac{k_{\mathrm{B}}T}{2Nq} \sum_\nu \int \frac{dk}{|G|} \ln\left[1\!+\!\exp\left({\frac{E_{\mathrm{F}}-E_{\nu k}}{k_{\mathrm{B}}T}}\right)\right]\nn\\
	+ (\bar{n}\!+\!1)E_{\mathrm{F}}- \frac{a}{2NL} E_{\mathrm{C}},\label{eq:fe}
\end{gather}
and identify the solution with the lowest $F$ as the most stable state.
Here $E_{\nu k}$ is the single particle energy in Eq.~\eqref{eq:hfeq},
and $E_{\mathrm{C}}$ is the constant energy shift obtained by Eq.~\eqref{eq:Ec}.
In Sec.~\ref{sec:tdw}, we present the most stable solution (ferromagnetic domain wall state, FDW) in doped $N=5$ GNR.
We will discuss relative stability of the different ordered states 
in Sec.~\ref{sec:stability}.

\begin{figure*}[t]
\begin{center}
	\includegraphics[width=150mm]{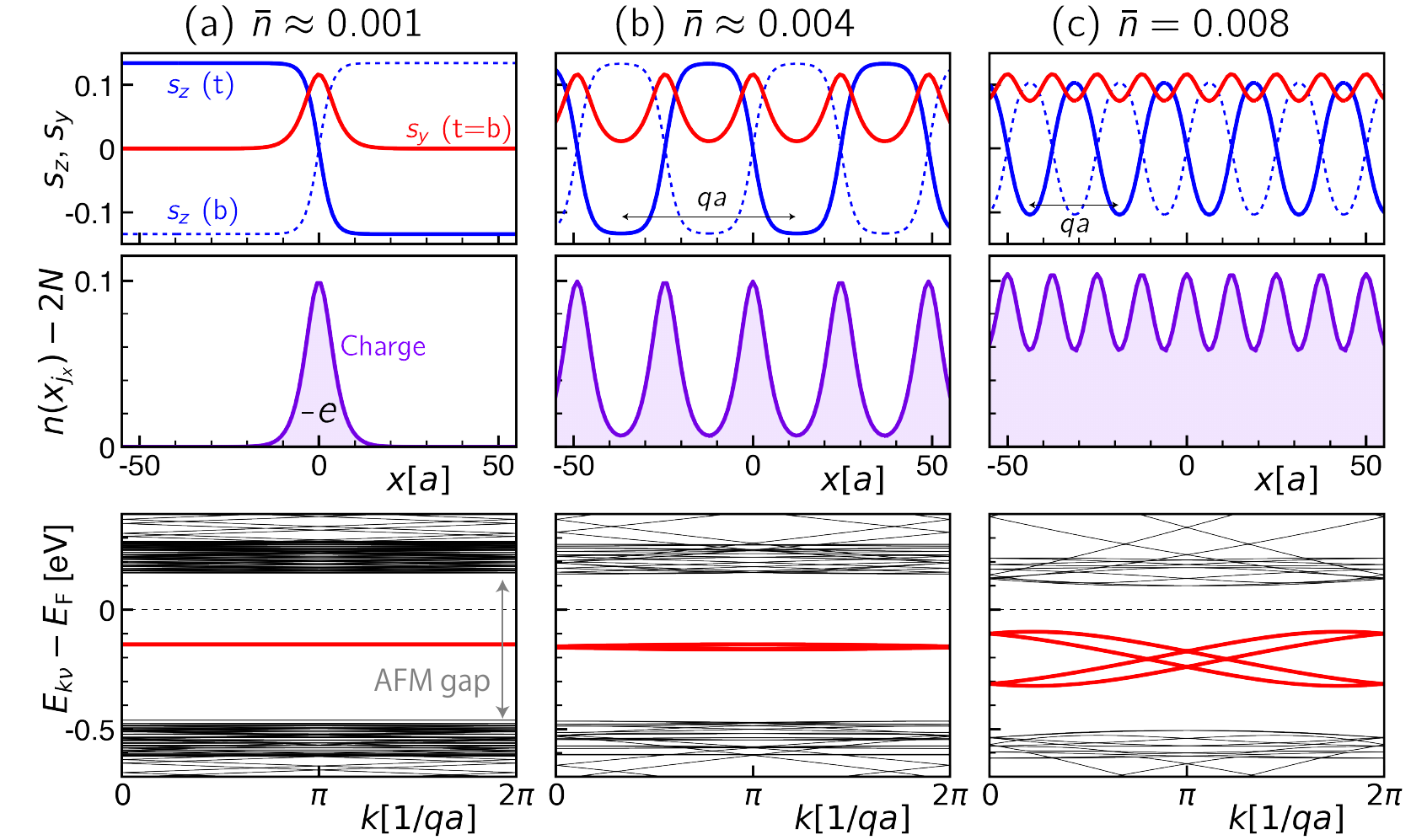}
	\caption{Ferromagnetic domain walls for $N=5$ ZGNR for several doped carrier densities 
	$\bar{n}\approx 0.001$ (a), $\bar{n}\approx0.004$ (b), and $\bar{n}=0.008$ (c) at $T=100$ K. 
	First row is the spin density as a function of $x_{j}$ at edge atoms.
	Solid (dashed) curves are for top (t) and bottom (b) edges and 
	blue (red) ones are for $s_z$ ($s_y$) components.
	The $s_y$ for the top/bottom edges are completely overlapped.
	Second row is the charge density in the ribbon unit cell $n(x_{m_x})=\sum_{m_y} n(\bm{r}_j)$. 
	Third row is the energy spectrum for obtained self-consistent order parameters.
	In-gap states are highlighted as red thick curves.
	}\label{fig:scsol}
\end{center}
\end{figure*}

\section{Ground State of Charged ZGNRs}\label{sec:tdw}

Figure \ref{fig:scsol} shows typical self-consistent solutions for width $N=5$ ZGNR 
with carrier dopings of $\bar{n}\approx$ 0.001, 0.004, and 0.008.
Here, we set the temperature to $T=100$ K, and we also confirmed the results remain unchanged in even lower temperatures.
The dependence on the ribbon width $N$ will be discussed in Sec.~\ref{sec:wdt}.
The first row of Fig.~\ref{fig:scsol} shows spatial profile of the local spin $\bm{s}(\bm{r}_j)=(0, s_y, s_z)$ 
at the top and bottom edges (indicated by solid and dashed curves, respectively) as a function of $x$.
The second row is the plot for the local charge density integrated over $y$.
The third row presents the energy spectrum obtained from Eq.~(\ref{eq:hfeq}).

In the lightly doped regime [e.g. $\bar{n}\approx0.001$ in Fig.~\ref{fig:scsol}(a)],  
the magnetic structure exhibits a series of AFM domains.
In the left (right) region away from $x=0$,
the spin at the top and bottom edges are uniformly polarized in $\pm \bm{z}$ ($\mp \bm{z}$), 
forming AFM ordered states as in the neutral ZGNRs \cite{fujita1996, son2006}.
On the domain boundary at $x\sim 0$, $s_z$ component is suppressed while $s_y$ comes out instead.
We note that the $s_y$ components at the top and bottom edges are ferromagnetically aligned.
The spatial profile of $s_y$ and $s_z$ in Fig.~\ref{fig:scsol} is schematically shown in Fig.~\ref{fig:schem}, where the spins at the two edges are continuously rotated with respect to $x$ axis, in opposite (clockwise and anti-clockwise) directions, resulting in the AFM-FM-AFM domain wall.
We refer to this domain structure as the ferromagnetic domain wall (FDW).
The same structure was first found in the previous work for $N=10$ \cite{sancho2017}. 

\begin{figure*}[t]
\begin{center}
	\includegraphics[width=170mm]{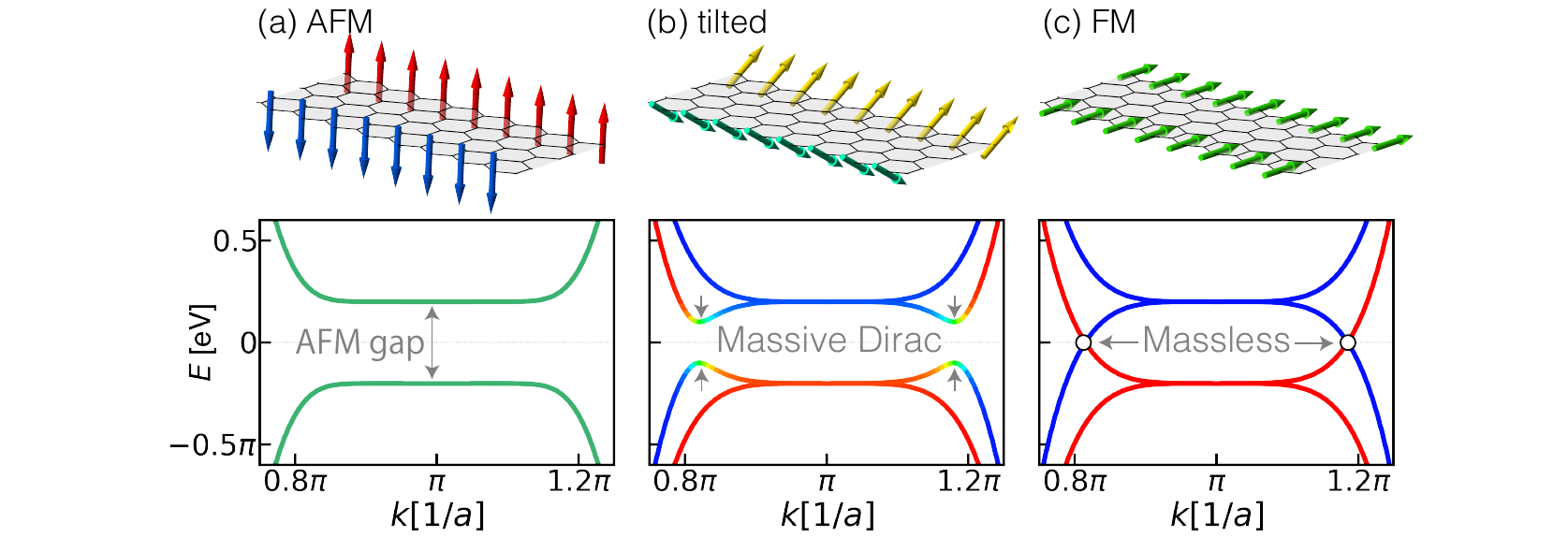}
	\caption{
	The typical magnetic orders at the edge atom (top panels) and 
	corresponding band structure (bottom).
	The order depicted in (a) creates the mass gap at the Dirac point in (c).
	}\label{fig:magnetic}
\end{center}
\end{figure*}

In the plot of the charge density [the middle panel in Fig.~\ref{fig:scsol}(a)], we see that the domain wall traps a doped charge of $-e$, which is associated with topological boundary modes.
In the energy spectrum [the bottom panel in Fig.~\ref{fig:scsol}(a)],
the boundary modes can be seen as highlighted red lines, which float in a bulk AFM energy gap [Fig.~\ref{fig:magnetic}(a)].
The emergence of the localized modes at the local FM region is consistent with the gapless spectrum in uniform FM order [Fig.~\ref{fig:magnetic}(c)].
Actually, the boundary modes in the FDW can be described as a topological localized state in a mass inversion 
of an effective Dirac model, this will be discussed in more detail in Sec.~\ref{sec:top}.

We can show that a single domain wall has two degenerate boundary modes, and the charge neutral point is located in the middle of the two (i.e., the boundary modes are half filled at the charge neutral point).
In the case of Fig.~\ref{fig:scsol}(a),
the doped carriers are accommodated in unoccupied boundary states, so that the Fermi energy comes to a large gap above the boundary states, and reduce the total energy.
Inversely speaking, a carrier-doped ZGNR automatically creates AFM domain boundaries such that the charged carriers exactly fill the in-gap topological boundary modes, and make the Fermi energy remain in the energy gap.

In increasing the carrier density, more and more domain walls emerge to absorb the doped carriers, and they come closer to each other as seen in Fig.~\ref{fig:scsol}(b).
In the heavily doped regime, the spin and charge spatial profile approach coexisting spin density wave (SDW) and charge density wave (CDW) 
with sinusoidal oscillation [Fig.~\ref{fig:scsol}(c)].
When the carrier density is further increased, the FM region eventually dominates and the system approaches a uniform FM phase.



\begin{figure}[b]
\begin{center}
	\includegraphics[width=85mm]{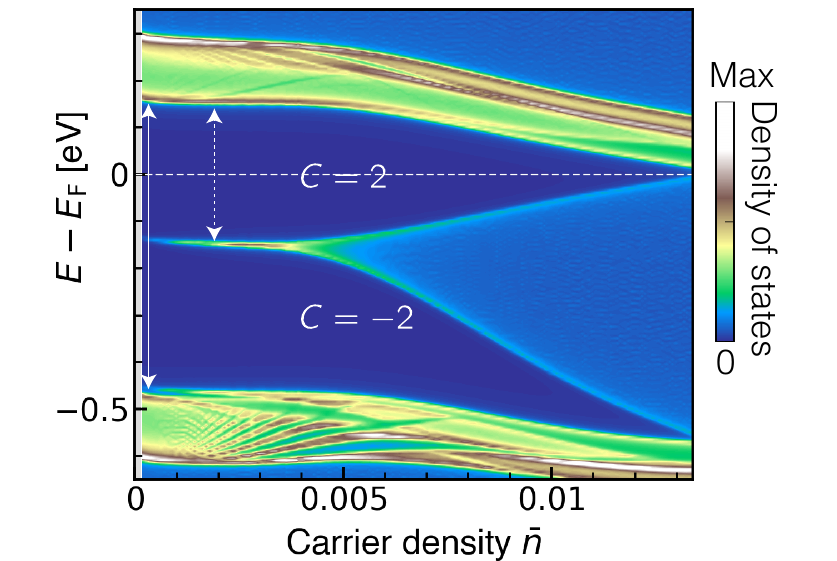}
	\caption{Spectral evolution of the domain wall state of GNR with width $N=5$ at $T=100$ K 
	as a function of doped carrier density $\bar{n}$.
	}\label{fig:butterfly}
\end{center}
\end{figure}

In the energy spectrum, the interference of the 
neighboring boundary modes leads 
to a finite energy width of in-gap bands as seen in the bottom panels of Fig.~\ref{fig:scsol}(c).
Figure~\ref{fig:butterfly} displays an evolution of the density of states as a function of the carrier density $\bar{n}>0$, 
which is calculated by 
\begin{align}
D(E)=\sum_{\nu} \int \frac{dk}{|G|}\frac{1}{\pi}\frac{\eta}{(E_{\nu k}-E)^2+\eta^2} 
\end{align}
with a smearing factor of $\eta=3$ meV.
We observe that the in-gap band is broadened in increasing $\bar{n}$, and it finally fills out the main energy gap. 
This corresponds to the metallic spectrum in uniform FM phase. [Fig.~\ref{fig:magnetic}(c)]

The diagram of Fig.~\ref{fig:butterfly} exhibits a series of mini gaps forming a complex structure reminiscent of the Hofstadter butterfly~\cite{hofstadter}. The Hofstadter butterfly is the energy spectrum in a two-dimensional (2D) lattice under magnetic field,
where the recursive pattern emerges as a function of the magnetic flux per a unit cell.
The same pattern also appears in the spectrum of a one-dimensional (1D)
doubly-periodic lattice such as Aubry-Andr\'e model~\cite{lang2012, kraus2012, kraus2012_2},
where the ratio of the two periods
corresponds to the magnetic flux in the Hosftadter's model.
The present system can be regarded as a 1D doubly-periodic system with two competing periods, the atomic lattice constant $a$ and the mean field superperiod $qa$,
and hence the Hofstadter-like structure appears as a function of $q \propto \bar{n}$ as seen in Fig.~\ref{fig:butterfly}.

Each mini gap of the Hofstadter butterfly is labeled by the (first) Chern number.
In the 2D lattice under the magnetic field, the number corresponds to the quantized Hall conductivity \cite{TKNN}, while that
in the 1D doubly-periodic system is related to an adiabatic charge pumping (Thouless pumping) associated with a relative slide of the two periodic potentials ~\cite{thouless1983, lang2012, kraus2012, kraus2012_2, fujimoto2020, zhang2020, su2020}.
For the present system, we calculate the Chern number by considering the quantized charge pump under a shift of the mean field $n^{\mu}(\bm{r}_j)$ 
with the Fermi energy fixed to each energy gap.
Considering a cylic shift $n^{\mu}(\bm{r}_j) \to n^{\mu}(\bm{r}_i +  q\bm{a} \delta)$ with shift parameter $\delta\, (0\leq \delta \leq 1)$, 
the Chern number is explicitly written as
\begin{equation}\label{eq:chern}
	C(\epsilon) = \sum_{E_{\nu k}<\epsilon} \int^{1}_{0} dk \int^{1}_0 d\delta (\partial_k A_{\delta}^{(\nu)} - \partial_{\delta}A_k^{(\nu)}).
\end{equation}
Here the wave number $k$ is normalized in units of $2\pi/qa$, 
$A_{\alpha}^{(\nu)}=i \sum_j u^\dag_{\nu k}(\bm{r}_j,\delta)[\partial_\alpha u_{\nu k}(\bm{r}_j,\delta)]$ is the Berry connection, 
and $u_{\nu k}(\bm{r}_j,\delta)$ is the solution Eq.~(\ref{eq:hfeq}) 
with the shifted potential $n^{\mu}(\bm{r}_j+ q\bm{a}\delta)$.
We find $C=\pm 2$ for the upper and lower main gaps, respectively, in Fig.~\ref{fig:butterfly}. 
The difference of the Chern numbers
between the two main gaps,
$\Delta C=4$, corresponds to the number of topological bound states in a single period.
Specifically, we have two domain walls per period and each domain wall accommodates two bound states (see, Sec.~\ref{sec:top}).
When the mean field is slid by a single period, therefore,
the number of pumped charges at any cross section is 4$e$.

\section{Topological Origin of magnetic domain walls}\label{sec:top}
The emergence of the domain-wall bound states can be understood by the following topological argument.
Below, we introduce a continuum model for the zigzag edge states in the presence of FM/AFM order, and demonstrate that the localized modes can be described as Jackiw-Rebbi states in a 1D Dirac system with a mass inversion.

In a non-interacting ZGNR with width $N$, the effective Hamiltonian of edge states around $k=\pi/a$ [Fig.~\ref{fig:honeycomb}(b)] can be written as
\begin{equation}\label{eq:ndirac}
h_{0} = \left[\begin{array}{cc} 
	0 & \tilde{k}^N \\  
	\tilde{k}^N & 0 
	\end{array}\right] \sigma_0 = \tilde{k}^N \tau_x \sigma_0.
\end{equation}
Here we take an energy unit with $t=1$ and define dimensionless wave number $\tilde{k}=k a -\pi$.
The first and the second elements in $2\times2$ Hamiltonian (Pauli matrix $\tau_x$) represent the top edge state $|\mathrm{t}\rangle$
and the bottom edge state $|\mathrm{b}\rangle$, respectively.
The derivation of Eq.~(\ref{eq:ndirac})
is provided in Appendix~\ref{sec:non-interact}.

The Hamiltonian of ZGNR with a magnetic domain wall in Fig.~\ref{fig:scsol} is expressed as
\begin{equation}\label{eq:h}
    h = h_0 + h_F + h_A,
\end{equation}
where
\begin{align}
\label{eq:potF}
& h_{\mathrm{F}} = f_y(x) \tau_0 \sigma_y,
\\
\label{eq:potA}
& h_{\mathrm{A}} = d_z(x) \tau_z \sigma_z,
\end{align}
are the FM and AFM exchange fields, respectively.
Here
$f_y(x) =(U/t) s_{y}(\bm{r}_j)$ and $d_z(x) =(U/t) s_{z}(\bm{r}_j)$ 
represent the spin densities at top edge sites
at $\bm{r}_j$ with $j=(m_x,1)$.
In the magnetic domain wall of Fig.~\ref{fig:schem}, 
the FM field dominates at the domain wall center, while the AFM field dominates far away from the domain wall.
In the following, we take the wave basis,
\begin{align}\begin{split}\label{eq:basis}
 &  |\psi_1\rangle=\tfrac{1}{2}( |\mathrm{t} \rangle + |\mathrm{b}\rangle)\otimes (\left|\uparrow\right>-i\left|\downarrow\right>), \\
 &  |\psi_2\rangle=\tfrac{1}{2}( |\mathrm{t} \rangle - |\mathrm{b}\rangle)\otimes (\left|\uparrow\right>+i\left|\downarrow\right>), \\
 & |\psi_3\rangle=\tfrac{1}{2}( |\mathrm{t} \rangle + |\mathrm{b}\rangle)\otimes (\left|\uparrow\right>+i\left|\downarrow\right>), \\
 &  |\psi_4\rangle=\tfrac{1}{2}( |\mathrm{t} \rangle - |\mathrm{b}\rangle)\otimes (\left|\uparrow\right>-i\left|\downarrow\right>), 
\end{split}\end{align}
where $|\mathrm{t} \rangle \pm |\mathrm{b}\rangle$ represent
bonding and anti-bonding states of top and bottom edges, respectively,
and $\left|\uparrow\right>\pm i\left|\downarrow\right>$
are spin states polarized to $\pm y$, respectively.
By using a unitary matrix $U_1=(|\psi_1\rangle, |\psi_2\rangle, |\psi_3\rangle, |\psi_4\rangle)$, the entire Hamiltonian $h=h_0+h_{\mathrm{F}}+h_{\mathrm{A}}$ is transformed to
\begin{equation}\label{eq:44}
     U_1^\dag h U_1 \!=\!
     \left[\begin{array}{cccc}
     \tilde{k}^N\!-\!f_y    & d_z & 0 & 0\\
     d_z & -\tilde{k}^N \!+\! f_y   & 0  & 0\\
     0 & 0 & \tilde{k}^N \!+\! f_y  &  d_z \\
     0 & 0 & d_z & -\tilde{k}^N\!-\!f_y
     \end{array}\right].
\end{equation}

When $f_y(x)$ and $d_z(x)$ changes slowly as a function of $x$, the local electronic structure at every single point can be approximately described by a uniform Hamiltonian with constant $f_y$ and $d_z$.
A domain-wall bound state emerges when the energy gap of the local band closes at a certain point in the domain wall.
Let us first consider a uniform ferromagnetic phase [Fig.~\ref{fig:magnetic}(c)]
corresponding to the domain-wall center.
The Hamiltonian is given by Eq.~\eqref{eq:44}
with $d_{z}=0$ and uniform $f_y>0$. 
It is clear that the band crossing occurs 
between the electron branch with $+y$-spin, $E_{+y}=|\tilde{k}|^N-f_y$ [red curve in Fig.~\ref{fig:magnetic}(c)], and hole branch with $- y$-spin, $E_{-y}=-|\tilde{k}|^N+f_y$ (blue). 
The crossing points of $E_{+y}=E_{-y}$ occur at   
\begin{equation} \label{eq:valley}
	 \tilde{k} = \pm f_y^{1/N} \equiv \pm k_{\mathrm{D}}.
\end{equation}
The energy bands near the crossing points  are regarded as 1D massless Dirac bands at two independent valleys $\pm k_{\mathrm{D}}$.

If we move away from the domain wall center,
spins are antiferromagnetically tilted (see Fig.~\ref{fig:schem}). The situation is described by introducing non-zero $d_z$ in Eq.~\eqref{eq:44}.
The off-diagonal elements $d_z$ immediately creates a mass gap in 1D Dirac cones as in Fig.~\ref{fig:magnetic}(b).
The effective 1D Dirac Hamiltonian near the band gap can be derived by expanding Eq.~\eqref{eq:44} around $\pm k_{\mathrm{D}}$.
For odd $N$, the low-energy states of the valley $+k_{\mathrm{D}}$ are associated with the upper left sector of Eq.~\eqref{eq:valley},
while the valley $-k_{\mathrm{D}}$ is associated with the lower left sector, as $k^N = -|k|^N$ for $k < 0$.
For even $N$, on the other hand, both of valleys $\pm k_{\mathrm{D}}$ are associated with the upper left sector.
By expanding the appropriate sector in Eq.~\eqref{eq:valley},
the 1D Dirac Hamiltonian at $ \pm k_{\mathrm{D}}$ is obtained as
\begin{equation}\label{eq:massivedk}
	h_{\pm}=\left[\begin{array}{cc}
		 v_{N}^\pm \tilde{k} & d_z \\
		 d_z & -v_{N}^\pm\tilde{k}  \\
	\end{array}\right],
\end{equation}
with the velocity
\begin{eqnarray}\label{eq:vdirac}
	v_{N}^{\pm} = (\pm 1)^{N+1} N f_y^{1-\frac{1}{N}}.
\end{eqnarray}
Here $\pm$ is the valley index for $\tilde{k} =\pm k_{\mathrm{D}}$. 

In the magnetic domain wall (Fig.~\ref{fig:schem}), 
the $d_z(x)$ continuously changes from positive to negative, 
and hence the mass in the Dirac electron is inverted at the domain wall center.
This causes zero-energy modes
called Jackiw-Rebbi bound states localized at the mass inversion point~\cite{jackiw}.
If we replace ${\tilde k} \rightarrow -i\partial_x$ and 
assume $d_z(x>0) < 0$ and $d_z(x<0) > 0$,
the zero-energy localized state can be explicitly written as
\begin{equation}\label{eq:j-r}
	\Phi_{0\pm}(x)
	=
	\exp\left( \frac{1}{|v_N^{\pm}|}
	\int_0^x d_z(x')dx' \right) 
	\left[\begin{array}{c}
		 1 \\
		 \mathrm{sgn}(v_{N}^{\pm}) i
	\end{array}\right].
\end{equation}
The bound states at the two valleys $\pm$ correspond to the doubly degenerate domain wall states in Fig.~\ref{fig:scsol}(a).

Besides the FDW considered above,
we can consider the AFM-DW structure shown in Fig.~\ref{fig:sols}(d),
where the spins of two edges rotate in the same direction retaining the local antiferromagnetic structure everywhere.
However, the state is found to be unstable
because the effective Hamiltonian with the AFM-DW exchange field 
$h_{\mathrm{ADW}} = h_0 + h_{\mathrm{A}} +  d_y(x)\sigma_y\tau_z$ is locally gapped everywhere,
giving no domain wall states to accommodate 
doped carriers.

\section{Thermodynamic stability of domain wall}\label{sec:stability}

\begin{figure}[t]
\begin{center}
	\includegraphics[width=90mm]{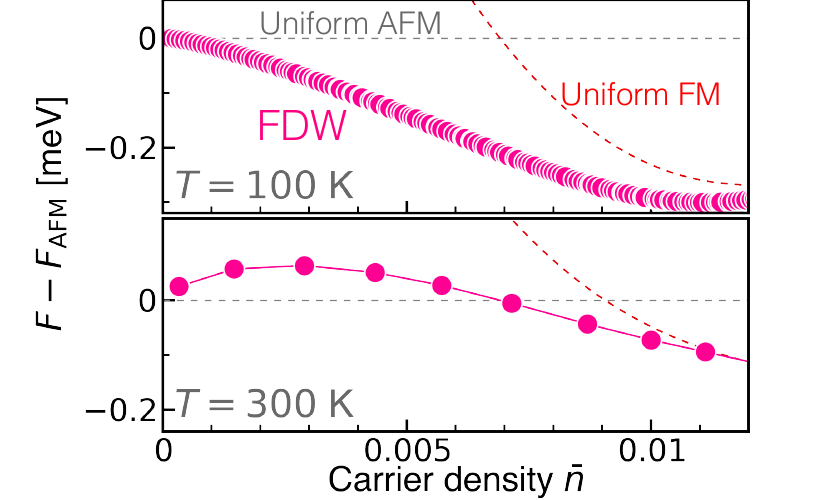}
	\caption{Free energy density of typical self-consistent solutions for width $N=5$ GNR as a function of doped carrier density $\bar{n}$.
	Filled symbol is for the FDW state and
	Gray (red) dashed curve for uniform AFM (FM).
	}\label{fig:few5}
\end{center}
\end{figure}

In this section, we discuss the thermodynamic stability of FDW state.
In Fig.~\ref{fig:few5}(a), 
we compare the relative free energies of FDW, uniform-AFM, and uniform-FM states [schematically shown in Fig.~\ref{fig:sols}] in $N=5$ GNR at $T=100$K,
by changing carrier density $\bar{n}$.
Here we set the free energy of uniform-AFM to zero for reference.
We actually see that the FDW phase 
has the lowest energy in wide range of the parameter.

Let us first consider the energetics of the uniform AFM and FM states.
At $\bar{n}=0$, the AFM is more stable than the FM, because the electronic spectrum of the AFM is gapped at the Fermi energy [Fig.~\ref{fig:magnetic}(a)], 
while that of the FM is gapless [Fig.~\ref{fig:magnetic}(c)].
In increasing $\bar{n}$, however, the energy difference is decreased and eventually the FM becomes more stable in $\bar{n} \gtrsim 0.007$.
This is because in the FM state, doped charges are accommodated in low-energy excited states near the charge neutrality point [Fig.~\ref{fig:magnetic}(c)],
whereas in the AFM,
doped carriers have to occupy high energy states above the gap, costing larger energy.

The free energy of FDW is smaller than that of uniform AFM and FM [Fig.~\ref{fig:few5}(a)], because
the doped carriers are accommodated to zero-energy domain-wall states [Fig.~\ref{fig:scsol}] with a small energy cost, while the rest of the system is gapped because of the local AFM order.
In other words, the FDW phase combines the characteristics of the AFM (domain) and FM (boundary) to reduce the total free energy.
In increasing $\bar{n}$, the free energy of FDW approaches that of the uniform FM,
since the FDW crosses over to the $y$-polarized SDW phase [Fig.~\ref{fig:scsol}(c)]
and eventually to the ferromagnetic state.

\begin{figure}[b]
\begin{center}
	\includegraphics[width=85mm]{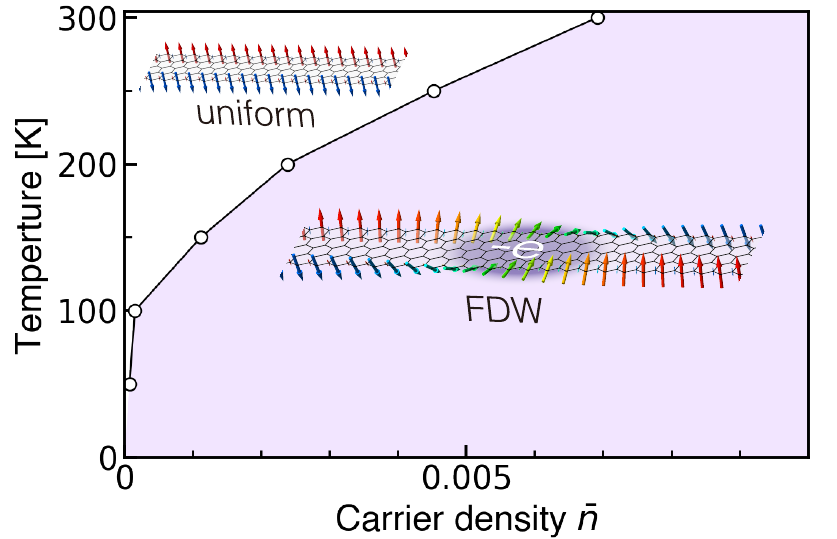}
	\caption{Phase diagram spanned by the temperature and doped carrier density. 
	In the shaded region, FDW structure is the ground state.
	The white region is the region where the metallic uniform state is stable.
	}\label{fig:phasediag}
\end{center}
\end{figure}

In $T=300$K, on the other hand,
we see that the uniform AFM phase has lower free energy than the FDW phase in the low-density region, $\bar{n} \lesssim 0.007$ [Fig.~\ref{fig:few5}].
The reason for this is that the uniform AFM state with carrier doping has a higher entropy $S$ due to a large degeneracy of single particle states above the gap,
compared to the FDW state where the dope carriers occupy exclusively the domain-wall bound states.
The decrease in the free energy by $-ST$ 
compensates the energy cost 
and stabilizes the uniform AFM phase at high temperature.
The phase diagram on the temperature and carrier density space is presented in Fig.~\ref{fig:phasediag}.

It should also be mentioned that the fluctuations of the domain wall distance, 
which is not included in the present mean field calculation, should also affect the thermodynamic stability of the FDW state.
As each domain wall traps a single electron, the walls repel each other by the Coulomb interaction with small screening due to one-dimensionality. 
For $\bar{n}\approx 0.004$ [see Fig.~\ref{fig:scsol} (b)], for instance, the spacing between domain walls is $\sim 60$\AA, where the bare Coulomb energy is estimated as $\sim 200$ meV. 
At the room temperature ($k_{\mathrm B} T\sim 30$ meV), therefore, the domain walls should be arranged in equal distance with a small fluctuation, similarly to a 1D Wigner crystal ~\cite{deshpande2008}.
The distance fluctuation should contribute to a configuration entropy, and make the FDW state more stable in a finite temperature. 
When the carrier density is even decreased, the domain walls are farther apart from each other, and domain walls move more freely to stabilize the FDW states.
When these effects are appropriately incorporated,
the FDW region in the phase diagram Fig.~\ref{fig:phasediag} would be expanded.
A further consideration on the quantum fluctuation and spin excitation~\cite{feldner2011, dutta2012, yoshioka2003, hikihara2003, karakonstantakis2013} would be presented elsewhere.

\section{Width dependence}\label{sec:wdt}
\begin{figure*}[t]
\begin{center}
	\includegraphics[width=170mm]{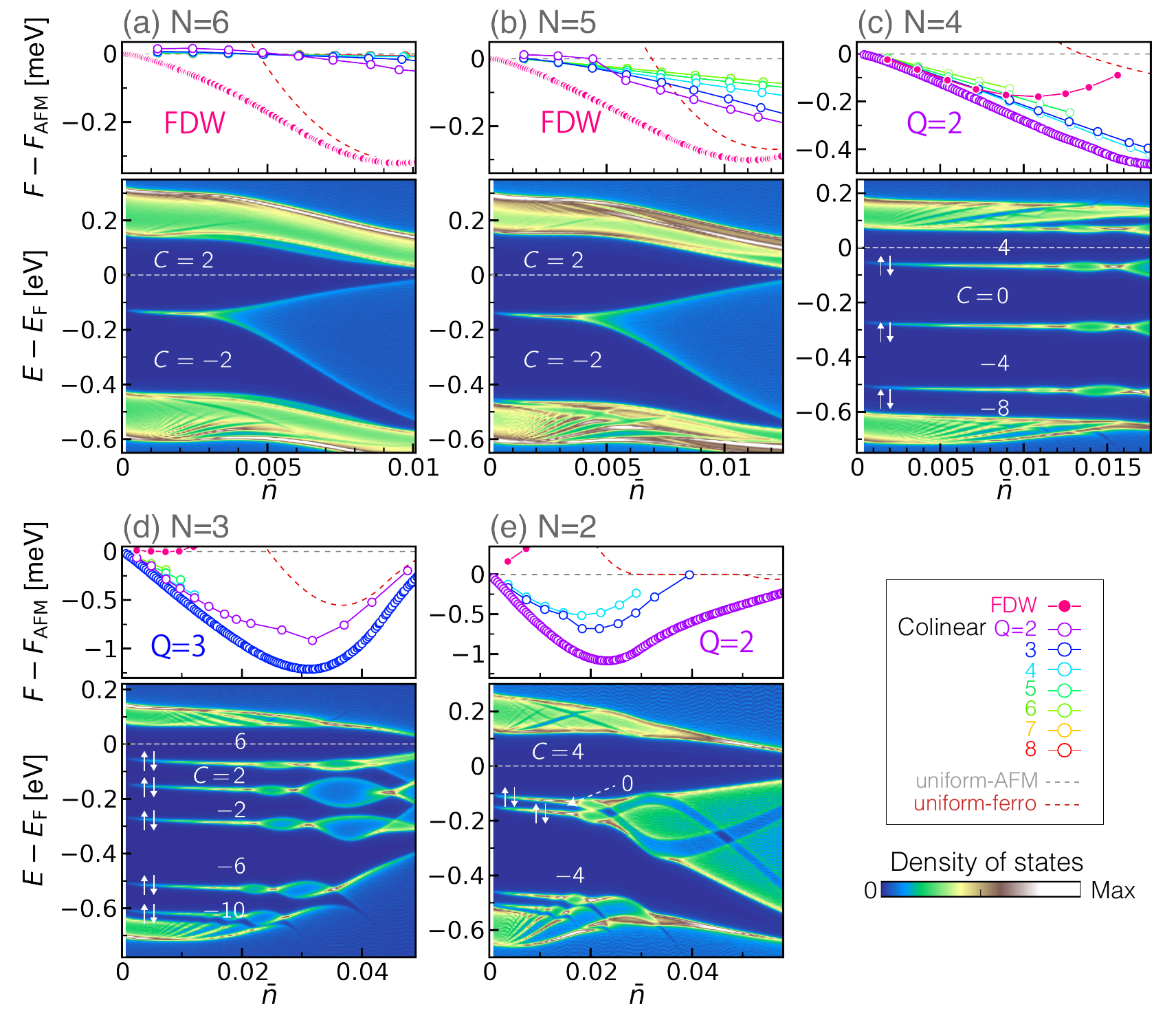}
	\caption{Free energy comparison and density of states for the most stable states for width $N=6$ (a), $N=5$ (b), $N=4$ (c), $N=3$ (d), and $N=2$ (e) and $T=100$ K.
	Color code of symbols in free energy plot are the same as in Fig.~\ref{fig:few5}. (b) includes a replot of Figs.~\ref{fig:butterfly} and~\ref{fig:few5}. 
	}\label{fig:wdt}
\end{center}
\end{figure*}

We investigate the domain wall formation in ZGNRs with different ribbon widths $N$.
Figure \ref{fig:wdt} summarizes
numerically obtained free energy and electronic spectra from $N=2$ to 6 plotted against the carrier density.
The ZGNRs of $N=5$ and 6 show qualitative similar properties argued in the previous sections, 
where the FDW is stable in a wide range of the doped regime.
We confirmed that the same trend persists in wider ribbons $N=7$ and 8.

In narrower ribbons with $N\le 4$, on the other hand, we find that a colinear domain wall phase [Fig.~\ref{fig:sols}(e)] emerges in doped regime, instead of the FDW phase.
It is also a domain boundary connecting opposite AFM regions, while different from the FDW in that 
spins are all parallel to $\pm z$ direction, and the spin polarization vanishes at the domain wall center.
The relative stability of the colinear phase over the FDW phase in narrow ribbons
can be understood as follows.
When we compare the band structures of uniform AFM and FM phases [Fig.~\ref{fig:magnetic}(a) and (c)], 
we understand that FM is less stable than AFM
due to the gap closing points,
while FM is more stable than the original non-magnetic state, 
because the central flat gapped region around $k=\pi$ reduces the free energy similarly to AFM.
When the width $N$ is decreased, however,
the flat region becomes narrower in $k$-space as shown in Fig.~\ref{fig:honeycomb}(b), so that the energy gain of the FM becomes smaller.
This effect makes the FDW state having local FM region unstable in narrow $N$'s.

The colinear domain wall also forms in-gap bound states as seen in Figs.~\ref{fig:wdt}(c)-(e),
which stabilize the domain wall by accommodating the doped charge in a similar way to FDW.
In contrast to the FDW where two degenerate bound states emerge at charge neutrality point, 
the colinear domain wall forms multiple in-gap levels,
and the number of the levels increases in increasing the width $N$.
This behavior can be qualitatively understood using the effective model of Eq.~\eqref{eq:h} with $h_F=0$, as presented in Appendix~\ref{sec:colinear}.
We also note that the in-gap states in the colinear phase 
are all doubly degenerate in spins.
This is because the spin $z$-component is a good quantum number 
in the colinear phase, and also the spin-up and spin-down sectors
are mirror inversions of each other, giving identical spectra.
The Fermi energy is located in one of the energy gaps which is chosen to minimize the free energy,
and it determines the number of trapped charges per a single domain wall, $Q$. 
Numerically checking the free energy as in the top panel of Fig.~\ref{fig:wdt} (c) and (d), we observe that single domain wall in $N=2$, $3$, $4$ trap $-2e$, $-3e$, and $-2e$ respectively.


As indicated in Figs.~\ref{fig:wdt}(c), (d), and (e), 
energy gaps in the colinear domain walls
are also labeled by Chern numbers in Eq.~\eqref{eq:chern} in the same manner as in the FDW. 
For the colinear domain wall in odd $N$ [Fig.~\ref{fig:wdt}(d)], 
each subgaps are labeled by the sequence of the Chern numbers $C= 2, 6, 10,\cdots$, 
while that in even $N$ [Fig.~\ref{fig:wdt}(c) and (e)] are $C=0,4,8,\cdots$. 
As the gap with $C=0$ corresponds to the charge neutral in the present system,
the difference of the sequences can be explained by
whether in-gap states emerge at the charge neutrality point or not, and it is related to the mirror winding number of the effective model~\eqref{eq:h}.
The detailed explanation for this is presented in the Appendix~\ref{sec:colinear}.
Also note that the difference of 4 between the Chern numbers of consecutive gaps is originating from the fact that each bound states has two-fold spin degeneracy and two the domain walls form a single period of the structure.


\section{Summary}\label{sec:summary}
We developed a theory of magnetically ordered states in carrier-doped zigzag graphene nanoribbons. 
Within the Hartree-Fock mean-field theory, 
we demonstrated that the carrier doping stabilizes
a magnetic structure with alternating AFM domains
separated by ferromagnetic domain walls,
where doped carriers are accommodated in topological bound states localized around domain walls.
We explained the topological origin of the domain wall bound states in terms of Jackiw-Rebbi arguments associated with 
the Dirac mass inversion.
With increasing the doped carrier, the spacing between domain walls is narrower, and the magnetic structure crosses over to the spin and charge density wave.
The energy spectrum shows the Hofstadter-like fractal spectral evolution as a function of the carrier density,
where minigaps are characterized by the Chern number associated with the quantized charge pump by the adiabatic shift of periodic domain walls. 
We systematically performed the calculation for GNRs with different widths,
and found that the ferromagnetic domain-wall phase emerges in wide ribbons of $N\geq 5$, while the colinear domain-wall phase arises $N\leq 4$.
As such, our finding provides new perspectives for studying the novel topological property of interacting ZGNR.

\begin{acknowledgements}
This work was supported in part by JSPS KAKENHI
Grant Number JP20K14415, JP21H05236, JP21H05232, JP20H01840,
JP20H00127, and 22J22312 and by JST CREST Grant Number JPMJCR20T3, Japan. 
The numerical calculations were partially performed on Yukawa21 at YITP
in Kyoto University.
\end{acknowledgements}

\appendix

\section{Non-interacting ZGNR and effective continuum model} \label{sec:non-interact}
\subsection{Energy spectrum of width-$N$ ZGNR}
We revisit the band structure and symmetric property of ZGNR without electron-electron interaction and spin degrees of freedom,
as a preparatory step for deriving the effective continuum model.
We use the Bloch basis
\begin{equation}
	| j_y, k \rangle = \sum_{j_x} e^{i k x_j} \psi_{j}^\dag | 0 \rangle
\end{equation}
where $\psi_{j}^\dag$ and $x_j$ are the spinless creation operator and $x$-coordinate of $p_z$ orbital at a site $j=(m_x,m_y)$ [definition of $m_x$ and $m_y$ is given in Fig.~\ref{fig:honeycomb}(a)].
In this basis the Bloch Hamiltonian of the ZGNR is described by the matrix,
\begin{eqnarray}\label{eq:zgnr}
	H_{0}(k) = 
	- t \left(\begin{array}{cccccc}
		0                  & C_{\frac{1}{2}} &   &  &  & \\
		C_{\frac{1}{2}} & 0                  & 1 &  &  & \\
		                   & 1                  & 0 & C_{\frac{1}{2}} &  \\
		                   &                    & C_{\frac{1}{2}} & 0 & 1 \\
		                   &                    &   & 1 & 0 & \\
		              &                    &   &   &  & \ddots\\
	\end{array}\right) \nn\\
	- t' \left(\begin{array}{cccccccc}
		0             & 0     & 0    & 1   &     &    & \\
		0             & 0     & C_1  & 0   & 0   &    & \\
		0             & C_1   & 0    & 0   & 0   & 1  & \\
		1             & 0     & 0    & 0   & C_1 & 0  &  \\
		              & 0     & 0    & C_1 & 0   & 0  &  \\
		              &       & 1    & 0   & 0   & 0  &  \\
		              &       &      &     &     &    & \ddots 
	\end{array}\right)
\end{eqnarray}
where we employ the shortened notation $C_{w}(k)=2 \cos(w k a)$.
Diagonalizing Eq.~(\ref{eq:zgnr}), we obtain the non-interacting energy spectrum for ZGNR with width $N$ shown in Fig.~\ref{fig:honeycomb}(b).

For later convenience, we block diagonalize the Hamiltonian Eq.~(\ref{eq:zgnr}),
by using the symmetry of the lattice structure. 
Clearly, ZGNRs with even width $N$ have a mirror reflection symmetry $M_{y}$ with respect to the $zx$ plane [see Fig.~\ref{fig:honeycomb}(a)].
Those with odd width $N$ have a glide mirror symmetry $\{M_{y}|\bm{a}/2\}$,
namely the combination of $M_{y}$ and half lattice translation by $\frac{\bm{a}}{2}$.
In the Bloch basis of Eq.~(\ref{eq:zgnr}), these two symmetries are described by the same matrix,
\begin{equation} \label{eq:gmat}
	G=\left(\begin{array}{cc}
	 0 & J_{N} \\
	 J_{N} & 0 
	\end{array}\right). 
\end{equation}
where $J_N$ is the $N\times N$ row-reversed identity matrix
\begin{equation}
	J_{N}=
	\left(\begin{array}{ccc}
	  &        & 1 \\
	  & \adots &  \\
	1 &        &  \\
	\end{array}\right). 
\end{equation}
The commutation relation $[H_{0}(k), G]=0$ allows us to block diagonalize Eq.~(\ref{eq:zgnr}) as 
\begin{eqnarray}\label{eq:block}
	U_{G}H_{0}(k)U_{G}^\dag = \diag (H_{0+}, \ H_{0-})
\end{eqnarray}
with a unitary matrix
\begin{equation}
	U_{G}=\frac{1}{\sqrt{2}}
	\left(\begin{array}{cc}
	  I_N  &  J_{N} \\
	 -I_N  &  J_{N} 
	\end{array}\right), 
\end{equation}
diagonalizing Eq.~(\ref{eq:gmat}). Here $I_N$ stands for the $N\times N$ identity matrix.
The block $H_{0 +}$ ($H_{0 -}$) in Eq.(\ref{eq:block}) is the Hamiltonian 
of the mirror or glide even (odd) states whose eigenenergy are plotted in purple (green) in Fig.~\ref{fig:honeycomb}(b).


\subsection{Derivation of the edge effective model} \label{sec:eff}
When we ignore further hopping $t'$ irrelevant in the wider ZGNR with $N\ge5$, 
the edge states can be described by continuum model with the significantly simple form Eq.~(\ref{eq:ndirac}) in the main text.
Let us derive this effective model.
First, we focus on the odd width $N$ and then apply the framework to even $N$ straightforwardly. 

Near the Brillouin zone boundary $ka = \pi +  {\tilde k}$ with small ${\tilde k}$ (normalized in the unit of $1/a$), 
we apply the linear expansion $2\cos(ka/2)\sim -{\tilde k}$ to glide-even block of Hamiltonian Eq.~(\ref{eq:block}) as 
\begin{equation}\label{eq:linear3}
	h_{0+} \sim
	\left( \begin{array}{ccccccc}
		0                & {\tilde k} &     \\
        {\tilde k}  & 0               & -1  \\
                         & -1              &  0  & {\tilde k}  \\
                         &                 & {\tilde k} &   0  \\
                         &                 &                 &      & \ddots \\
                         &                 &                 &      &        &  0 & -1 \\
                         &                 &                 &      &        & -1 & {\tilde k}  
	\end{array}\right)
\end{equation}
for the ZGNR with odd $N$. Here $h_{0+}=H_{0+}/t$ is a dimensionless Hamiltonian and we set $t'=0$ for the simplicity.
At the Brilloiuin zone boundary ${\tilde k}=0$, 
Eq.~(\ref{eq:linear3}) has zero energy solution $\epsilon_{\nu, +, k}=0$ with the wave function 
$\vec{u}_{\nu}(\pi/a)=(1,0,\cdots,0)^T$. 
Slightly away from $\tilde{k}\sim 0$, we can take into account the effect of finite ${\tilde k}$ with the second order perturbation theory
\begin{equation}\label{eq:perturbp}
	h_{0+}^{\mathrm{eff}} = -V^\dag (h_{0+}')^{-1} V
\end{equation}
where $V^\dag(k) = ({\tilde k}, 0, \cdots , 0)$ corresponds to $1\times (N-1)$ upper right block of Eq.~(\ref{eq:linear3}), and $h_{0+}'(k)$ to $(N-1)\times (N-1)$ lower right block.
In addition, one can analytically obtain the inverse of $h_{0+}'$, 
\begin{align}\label{eq:inv3}
	&(h_{0+}')^{-1} =  \nn\\
	&-\left[ \begin{array}{lllllllll}
	{\tilde k}^{N-2}  &  1      & {\tilde k}^{N-3} & {\tilde k}^1     & {\tilde k}^{N-4} & {\tilde k}^2      &\cdots & {\tilde k}^{\frac{N-1}{2}-1} \\
	\vdots  &  0 &  0      &  0      &  0      &  0       &\cdots &  0 \\
	  & \vdots   & {\tilde k}^{N-4} &  1      & {\tilde k}^{N-5} & {\tilde k}^1      &\cdots & {\tilde k}^{\frac{N-1}{2}-2}  \\
	  &    & \vdots &  0      &  0      &  0       &\cdots &  0  \\
	  &    &  & \vdots  & {\tilde k}^{N-6} &  1       &\cdots & {\tilde k}^{\frac{N-1}{2}-3}\\
	  &    &  &   & \vdots  &    0 &\cdots & 0 \\
	  &    &  &   &   & \vdots   &\ddots &\vdots
	\end{array}\right],
\end{align}
where the lower-triangular elements are symmetric to the upper.
Note that only the upper left-most  corner element of Eq.~(\ref{eq:inv3}) is relevant to Eq.~(\ref{eq:perturbp}).
As the result, we obtain effective Hamiltonian for glide even sector,
\begin{equation}
    h_{0+}^{\mathrm{eff}} = {\tilde k}^N
\end{equation}

In addition, those for glide odd sector is straightforwardly derived within the same process. 
Indeed, the glide-odd block of Eq.~(\ref{eq:block}) is related to that of glide even sector Eq.~(\ref{eq:perturbp}) as
\begin{equation}\label{eq:perterbm}
	h_{0-} = - U_{\mathrm{ex}}h_{0+}U_{\mathrm{ex}}^{\dag}
\end{equation}
with an anti-diagonal exchange operator
\begin{align}
	U_{\mathrm{ex}} = 
	\left(\begin{array}{ccccc}
	&&&&-1 \\
	&&&1&  \\
	&&-1&& \\
	&1&&&  \\
	\adots&&&  \\
	\end{array}\right)
\end{align}
The unitary equivalence Eq.~(\ref{eq:perterbm}) leads us to the effective model
\begin{equation}
	h_{0-}^{\mathrm{eff}} = -{\tilde k}^N.
\end{equation}
Finally, by changing the basis to the original Bloch basis, 
we reach the effective model
\begin{equation}\label{eq:effap}
	h_0=e^{-i\tau_y \frac{\pi}{4}}
	\left(\begin{array}{cc}
	h_{0+}^{\mathrm{eff}} & 0 \\
	0 & h_{0-}^{\mathrm{eff}}\end{array}\right) e^{i\tau_y \frac{\pi}{4}} = 
	\left(\begin{array}{cc}
	0 & {\tilde k}^N \\
	{\tilde k}^N & 0
	\end{array}\right).
\end{equation}
where $\tau_y$ is a 2$\times$2 Pauli matrix.

The procedure for the odd width $N$ discussed above is applicable to the even $N$.
Only differences are in a small part of matrix elements of Eq.~(\ref{eq:linear3}).
For the even $N$ its matrix form is replaced as 
\begin{align}\label{eq:linear}
	h_{0+} =
	\left( \begin{array}{ccccccc}
		0                & {\tilde k}             &      \\
        {\tilde k}	             & 0               &  -1  \\
                         & -1              &  0              & {\tilde k}  \\
                         &                 & {\tilde k}             &   0  \\
                         &                 &                 &      & \ddots \\
                         &                 &                 &      &        & 0    & {\tilde k} \\
                         &                 &                 &      &        & {\tilde k}  & -1
	\end{array}\right),
\end{align}
Because inverse of lower right $(N-1)\times(N-1)$ block of Eq.~\eqref{eq:linear} is totally the same as Eq.~(\ref{eq:inv3}), 
the effective Hamiltonian for the even width $N$ is also given by Eq.~\eqref{eq:effap}.

\section{Bound states in colinear domain walls}~\label{sec:colinear}
We demonstrate that the colinear domain walls [Fig.~\ref{fig:sols}(e)] form in-gap bound states
and topologically characterize those in the ribbons with width $N$.


\subsection{Effective model description}

Electronic structure of the colinear domain walls is
described by the continuum model Eq.~(\ref{eq:ndirac}) 
and AFM exchange field Eq.~(\ref{eq:potA}).
The total Hamiltonian has block-diagonal form $h_0+h_\mathrm{A}=\mathrm{diag}[h_{N+},h_{N-}]$ with spin-up and -down sectors,
\begin{equation}\label{eq:coldirac}
	h_{N s} = 
	\left[\begin{array}{cc}
		s d_z & {{\tilde k}}^N \\
		{{\tilde k}}^N  & -s d_z
	\end{array}\right]=\tilde{k}^{N}\tau_x + s d_z\tau_z, 
\end{equation}
where $s=\pm$ stands for the spin up and down, and the basis of the $2\times 2$ matrix 
are the top and bottom edges. 
Here, spin up and down correspond to the states with even and odd parity under the mirror reflection with respect to $xy$-plane, respectively.
The Hamiltonian Eq.~\eqref{eq:coldirac} is the one-dimensional massive Dirac model with $N$-th order kinetic term ${\tilde k}^N$.
The mass $d_z$ changes the sign at the domain wall.
Hereafter in this section we only focus on the mirror even (spin up) state,
because the mirror odd (down-spin) Hamiltonian is unitary equivalent to that of even (up),  
$h_{N -} = \tau_x h_{N +}\tau_x$. 

\begin{figure}[t]
\begin{center}
	\includegraphics[width=75mm]{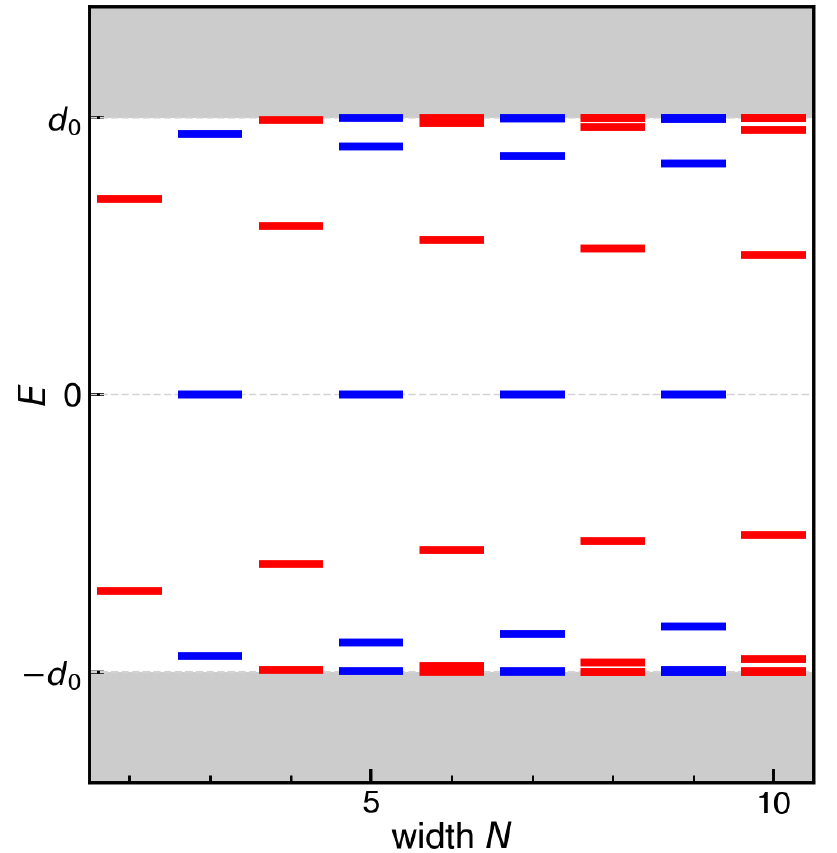}
	\caption{Numerically obtained bound state for colinear domain wall with width $N$.
	Gray area indicates the spectra of propagating modes with antiferromagnetic gap $2d_0$.
	}\label{fig:nthjr}
\end{center}
\end{figure}

We can obtain the bound states by expanding Jackiw-Rebbi argument of the $k$-linear Dirac model~\cite{jackiw} 
to the $N$-th order model Eq.~(\ref{eq:coldirac}). 
When spin polarization $d_z$ is spatially uniform $d_z(x)=d_0$, we have propagating modes with energy
\begin{eqnarray}\label{eq:egnprop}
	E=\pm \sqrt{d_0^2+{\tilde k}^{2N}} 
\end{eqnarray}
with the gap $-|d_0|<E<|d_0|$.
In the real space basis, their eigen wave function is
\begin{eqnarray}
	\psi_{E,k}(x) = 
	\left(\begin{array}{c} 
	{\tilde k}^N  \\
	E-d_0
	\end{array}\right) e^{i{\tilde k} x}.
\end{eqnarray}
Let us consider the model of sharp colinear domain wall 
\begin{eqnarray}~\label{eq:step}
	d_z(x) = d_0 \theta(x).
\end{eqnarray}
where $\theta(x)$ is Heaviside step function $\theta(x)=1$ for $x>0$ and $\theta(x)=-1$ for $x<0$.
Extending the momentum in Eq.~(\ref{eq:egnprop}) to the complex plane ${\tilde k}=\kappa e^{i\chi}$
we obtain the in-gap evanescent mode with the energy 
\begin{eqnarray}\label{eq:egneva}
	E=\pm \sqrt{d_0^2-\kappa^{2N}}.
\end{eqnarray}
Here, because the energy Eq.~(\ref{eq:egneva}) should be real and smaller than the gap of the propagating modes,
the complex momentum is required to be ${\tilde k}=\pm k_{n}= \pm \kappa e^{i(2n-1)\pi/2N}$ with $n=1,2, \cdots N$.
The wave function of these evanescent modes is
\begin{eqnarray}
	\varphi_{E,n}^{(\mathrm{L/R})}(x) = 
	\left(\begin{array}{c} 
		\mp k_n^N  \\
		E\pm d_0
	\end{array}\right) e^{\mp ik_n x}.
\end{eqnarray}
where upper (lower) sign is for the index L (R) in left hand side, indicating the decaying modes in $x<0$ ($x>0$).
The wave function of the bound state in the two sides of the domain wall is given by the linear combination of these modes, 
\begin{equation}
	\Phi^{\mathrm{(L/R)}}_{E}(x) = \sum_{n} C_{\mathrm{L/R},n}(E)\varphi_{E,n}^{(\mathrm{L/R})}(x) 
\end{equation}
To avoid the divergence of the $N$-th order derivative in the Hamiltonian Eq.~\eqref{eq:coldirac},
we require smooth connection of $\Phi^{\mathrm{(L)}}_{E}(x)$ and $\Phi^{\mathrm{(R)}}_{E}(x)$ at $x=0$
\begin{eqnarray}\label{eq:cond}
	\big[\partial_x^{n} \Phi^{\mathrm{(L)}}_{E}(x)\big]_{x=0} = \big[\partial_x^{n} \Phi^{\mathrm{(R)}}_{E}(x)\big]_{x=0} 
\end{eqnarray}
with $n=0, 1, \cdots, N-1$. 
As a result, Eq.~(\ref{eq:cond}) falls into the homogeneous linear equation for $\bm{C}(E) = (C_{\mathrm{L},1},\cdots C_{\mathrm{L},N}, C_{\mathrm{R},1},\cdots C_{\mathrm{R},N})$ 
\begin{equation}
	A(E) \bm{C}(E) = 0
\end{equation}
Numerically solving $\det[A(E)]=0$, we obtain quantized boundary modes $E=E_{\nu}$ as shown in Fig.~\ref{fig:nthjr}. 
Importantly, for width $N$ graphene nanoribbon, we have $2N$ bound states in total ($N$ states per a spin).

\subsection{Even-odd effect and winding number}
In Fig.~\ref{fig:nthjr}, we see the even-odd effects on the ribbon width $N$:
ribbon with odd $N$ has zero energy bound states at the charge neutral point
while that with even $N$ has an energy gap there.
This difference can be understood in terms of the winding number for the effective Hamiltohian Eq.~\eqref{eq:coldirac}.
The key point is that the Hamiltonian Eq.~\eqref{eq:coldirac} has a chiral symmetry
\begin{equation} \label{eq:chiral}
    \Gamma h_{N \pm} \Gamma^\dag=-h_{N \pm} , \hbox{ with } \Gamma= \tau_y.
\end{equation}
Taking the basis diagonalizing the chiral operator $\Gamma$,
we can transform the Hamiltonian to anti-diagonal form,
\begin{equation}
    U h_{N \pm} U^\dag =
    \left[\begin{array}{cc}
        0 & {{\tilde k}}^N \pm i d_z  \\
        {{\tilde k}}^N\mp id_z  & 0
    \end{array}\right],
\end{equation}
by a unitary matrix $U=e^{i\pi \tau_x/4}$.
For a uniform system with constant $d_z$,
we can define the one-dimensional winding number~\cite{sato2011} in each mirror sector,
\begin{align}
    w_{N \pm} (d_z,f_y)&= \frac{1}{4\pi i} \int d \tilde{k}\,{\mathrm{tr} \Big\{\Gamma [h_{N\pm}(\tilde{k})]^{-1}\partial_{\tilde k} h_{N\pm}(\tilde{k})\Big\}} \nn \\
    &= \frac{1}{2\pi i} \int d \tilde{k} \partial_{\tilde k} \ln ({{\tilde k}}^N \pm i d_z)
\end{align}
The independent topological numbers of the system are total winding number $W_{\mathrm{tot}}=w_{N+}+w_{N-}$ and mirror winding number $W_{\mathrm{m}}= w_{N+}-w_{N-}$.
For the present system, the total winding number is $W_{\mathrm{tot}}=0$ for any width $N$, while the mirror winding number is
\begin{align}
W_{\mathrm{m}} =
    \left\{\begin{array}{l}
    \mathrm{sign}(d_z) \quad \hbox{ for odd } N\\
    0  \quad \hbox{ for even } N
    \end{array}\right.
\end{align}
For the odd $N$ ribbons, the difference of the mirror winding number by 2 between $d_z>0$ and $d_z<0$ regions
topologically protects the twofold degenerate zero modes as in Fig.~\ref{fig:nthjr}.

In the self-consistent mean field Hamiltonian Eq.~\eqref{eq:hfeq}
the local charge modulation, or say CDW, breaks the chiral symmetry.
However, as long as the perturbation is small enough and the level inversion of in-gap states does not occur,
the even-odd effect about the bound states at charge neutral point
remains unchanged.
This is why the series of the possible chern number for even width include $C=0$ (charge neutral gap) [Figs.~\ref{fig:wdt}(c) (e)]
while those of odd width do not [Figs.~\ref{fig:wdt}(d)].

We see in Fig.~\ref{fig:wdt}(e) that the level spacing at the charge neutral gap with $C=0$
is much smaller than $E=0$ gap of $N=2$ in Fig.~\ref{fig:nthjr}.
This is due to the third nearest neighbor hopping $t'$ in Eq.~\eqref{eq:h0}
supporting band overlapping and crossing of conduction and valence band 
in the non-interacting band structure Fig.~\ref{fig:honeycomb}(b).
Two crossing in $N=2$ [Fig.~\ref{fig:honeycomb}(b)] 
are regarded as massless Dirac points and AFM order opens mass gap there.
As the mass inversion occurs at the colinear domain wall, there appear two Jackiw-Rebbi zero modes,
which correspond to two energetically close bound states in Fig.~\ref{fig:wdt}(e).
The small splitting of in-gap level in Fig.~\ref{fig:wdt}(e) are caused by the inter valley coupling. 
Therefore, with reducing $t'$ and mixing valleys,
the two levels strongly split as in Fig.~\ref{fig:nthjr}.

It should be noted that the sharp domain wall limit Eq.~\eqref{eq:step} also enhances level spacing in Fig.~\ref{fig:nthjr} for general $N$.
When the size of the domain wall is finite,  e.g. by replacing step function $\theta(x)$ in Eq.~\eqref{eq:step} to $\tanh(x/\xi)$, 
the domain wall behaves as a potential well with the size $\xi$.
In general, with increasing the size of the well $\xi$, the level spacing of the bound states tends to be smaller 
and additional bound states come out at the gap edges. 
This effect is taken into account in Fig.~\ref{fig:wdt}, 
where the number of the bound states are larger than $N$ 
in contrast to Fig.~\ref{fig:nthjr}.

\section{Optical absorption spectra}
Optical absorption spectroscopy, which is a basic tool to observe the electronic structure of the graphene-based systems,
can probe the topological bound states and characteristic spectral evolution in Fig.~\ref{fig:butterfly}.
Applying the linear response theory to the numerical solution of Eq.~(\ref{eq:hfeq}), 
we here evaluate the dynamical conductivity of the doped ZGNR,
\begin{eqnarray}
	\sigma_{\alpha\alpha}(\omega) = \frac{e^2\hbar}{iS} \sum_{\nu,\nu'} \int \frac{dk}{|G|}\frac{f(E_{\nu k})\!-\!f(E_{\nu' k})}{E_{\nu k}\!-\!E_{\nu' k}}\nn\\
	\times \frac{\left|\sum_{i,j}'u_{\nu k}^\dag(\bm{r}_i)v^{ij}_{\alpha} u_{\nu' k}(\bm{r}_j)\right|^2}{E_{\nu k}\!-\!E_{\nu' k}\!+\!\hbar{\omega}\!+\!i\eta}
\end{eqnarray}

\begin{figure}[b]
\begin{center}
	\includegraphics[width=85mm]{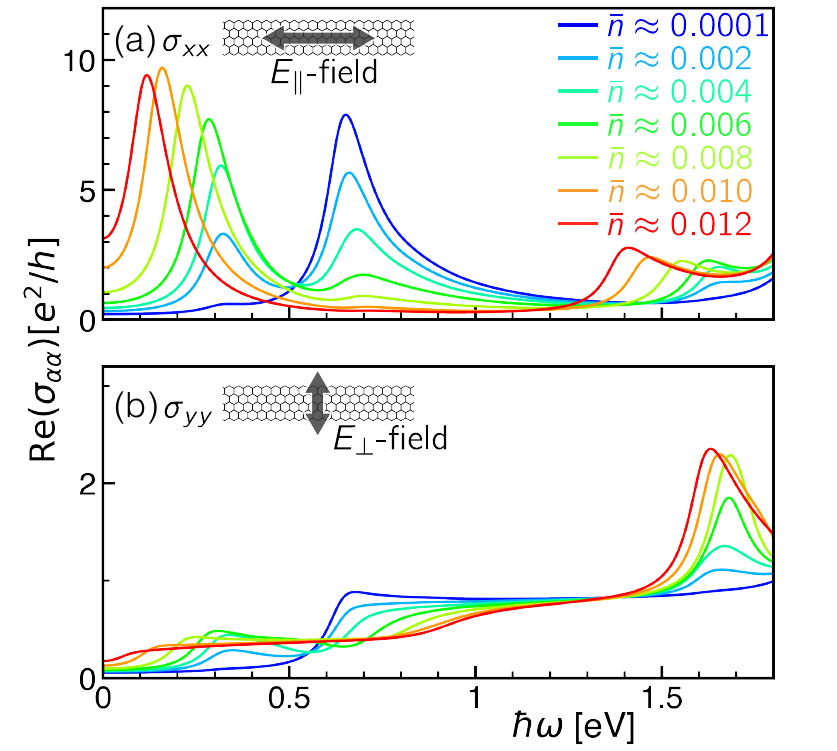}
	\caption{Dynamical conductivity driven by the parallel $\sigma_{xx}$(a) and perpendicular $\sigma_{yy}$(b) ac electric field. 
	Color code indicate the amount of the doped charge per a lattice point.
	}\label{fig:optical}
\end{center}
\end{figure}
 
where $S=(\sqrt{3}/2)Nqa^2$ is the size of the two dimensional super unitcell, $v^{ij}_\alpha$ is the velocity operator and $\alpha=x,(y)$ is parallel (perpendicular) to the ribbon.
The $x$-component of velocity is given by $v_{x}^{ij}=\hbar^{-1} \partial_k H_{k}^{ij}$ with the kinetic term $H_{k}^{ij}$ and 
$y$-component is $v_{y}^{ij}= i\hbar^{-1} (y^iH^{ij}_{k}-H^{ij}_k y^j)$. 
The real part of $\sigma_{\alpha\alpha}$ is proportional to the absorption probability 
of the light linearly polarized along $\alpha=x,y$. 
The obtained results are shown in Fig.~\ref{fig:optical}. 
Absorption of $y$ component of the light [Fig.~\ref{fig:optical}(b)] is smaller than that of $x$ component [Fig.~\ref{fig:optical}(a)],
since the edge states, relevant for the lower frequency spectra displayed in Fig.~\ref{fig:optical}, 
are confined along $x$ direction hardly move along $y$ direction.

The emergence of the topological bound states affects the dynamical conductivity $\sigma_{xx}$ along the ribbon [Fig.~\ref{fig:optical}(a)].
For the extremely light doping [blue curve in Fig.~\ref{fig:optical}(a)], the main peak appears at $\hbar \omega\approx0.6$ eV, 
corresponding to the excitation energy between the AFM gap edges (solid arrow in Fig.~\ref{fig:butterfly}). 
The bound states give rise to just a shoulder structure at $\hbar\omega\approx 0.3$ eV due to their diluteness.
When the doped carrier slightly increases (light-blue curve in Fig.~\ref{fig:optical}), we see the development of the lower frequency peak in $\sigma_{xx}$
caused by the excitation from the occupied bound states to the unoccupied conduction band (dashed arrow in Fig.~\ref{fig:butterfly}).
At the same time the peak at $\hbar \omega\approx0.6$ eV is continuously suppressed, 
since the gap-edge states contribute to form the bound states.
Due to the shrinkage of the gap around the Fermi energy (see Fig.~\ref{fig:butterfly}),
the lower energy peak continuously shifts to the zero frequency.
In particular, the double-peak structure in Fig.~\ref{fig:optical}(a) 
from the infrared to microwave region is the typical signal 
indicating the coexistence of the AFM gap and the topological bound states.

\bibliography{reference}
\end{document}